\documentclass[a4paper,11pt]{article}

\usepackage{aas_macros}
\usepackage{physics}

\usepackage[margin=1in]{geometry}

\usepackage{graphicx}	% Including figure files
\usepackage{amsmath}	% Advanced maths commands
\usepackage{amssymb}	% Extra maths symbols
\usepackage{hyperref}
\usepackage{authblk}
\usepackage{multirow}

\usepackage{rotating}

\usepackage[dvipsnames]{xcolor}

\hypersetup{
    colorlinks=true,
    urlcolor=ForestGreen,
    linkcolor=blue,
    citecolor=purple}%magenta}

\newcommand{\smica}{\texttt{SMICA}}

\newcommand{\planck}{\textit{Planck}}
\newcommand{\nside}{$N_{\rm side}$}
\newcommand{\nsidesq}{$N_{\rm side}^2$}
\newcommand{\npix}{$N_{\rm pix}$}
\newcommand{\healpix}{\texttt{HEALPix}}
\newcommand{\healpy}{\texttt{Healpy}}

\title{\bfseries Probing foreground residuals in cleaned CMB temperature maps from \planck}

\author[1]{Sanjeet K. Patel}%\thanks{sanjeetkumarpatel.rs.phy18@itbhu.ac.in}}%\thanks{sanjeetkumarpatel.rs.phy18@itbhu.ac.in}}
\author[1]{Pavan K. Aluri\thanks{pavanaluri.phy@itbhu.ac.in}}%\orcidlink{https://orcid.org/0000-0002-4333-9819}}
\author[2]{Pranati K. Rath}%\thanks{pranati13@gmail.com}}
\author[3]{Pramoda K. Samal\thanks{pksamal@gmail.com}}%\orcidlink{https://orcid.org/0000-0003-1490-4679}}
\affil[1]{Dept. of Physics, Indian Institute of Technology (BHU), Varanasi - 221005, India}
\affil[2]{Dept. of Physics, Khallikote Unitary University, Berhampur, Odisha - 760001, India}
\affil[3]{School of Physics, Gangadhar Meher University, Sambalpur, Odisha - 768004, India}

\date{\today}

\begin{document}

\maketitle

%Abstract of the paper
\begin{abstract}
Maps of cosmic microwave background (CMB) are extracted from multi-frequency observations using a variety of cleaning procedures. However, in regions of strong microwave emission, particularly in the galactic plane from our own galaxy Milky Way and some extended or point sources, the recovered CMB signal is not reliable. Thus, a galactic mask is provided along with the \emph{cleaned} CMB sky for use with that CMB map which excises sky regions that may still be potentially contaminated even after cleaning. So, to avoid bias in our inferences, we impose such a foreground mask. In this paper, we analyze a cleaned CMB map from \planck\ PR4 to probe for any foreground residuals that may still be present \emph{outside} the galactic mask where the derived CMB sky is considered clean. To that end, we employ a local cross-correlation coefficient statistic where we cross-correlate widely used foreground templates that trace galactic synchrotron, free-free, and thermal dust emission from our galaxy with the cleaned CMB sky. Using simulations, we find that few regions of the derived CMB sky are still contaminated and have to be omitted. Based on this study, we derived a mask that could be used in conjugation with the standard mask to further improve the purpose of galactic masks.
\end{abstract}

%\begin{keywords}
%Cosmic Microwave Background -- foregrounds
%\end{keywords}

%%%%%%%%%%%%%%%%%%%%%%%%%%%%%%%%%%%%%%%%%%%%%%%%%%

%%%%%%%%%%%%%%%%% BODY OF PAPER %%%%%%%%%%%%%%%%%%

\section{Introduction}
The cosmic microwave background (CMB) has been pivotal in establishing the current standard model of cosmology. So far, precise measurements of CMB temperature anisotropies have been made from a variety of ground-based and space-bourne missions mapping up to very small angular scales. Much of the cosmological constraints derived from full-sky missions are due to these precise temperature anisotropies, while polarized-sky measurements are affected by instrumental noise~\cite{Bennett2013finalmaps,Planck2018overview}. Full-sky polarization measurements with similar signal-to-noise ratio are the target of next-generation CMB experiments~\cite{PIXIE_Kogut_2011,PICO_2019,CMB_Bharat_Adak_2022,SimonsObs2019,CMBS4ScienceCase2019,GroundBIRD2020}. Together with the cosmic signal of our interest, viz. CMB, the detectors also register any emission irrespective of its origin falling in its designed frequency range of detection.

In order to extract cosmological information from the CMB, we have to separate the astrophysical \emph{foregrounds} from the cosmic microwave \emph{background}. Many techniques were developed to do this \emph{cleaning} of observed/raw maps measured at different frequencies, i.e. to remove foreground contamination from the observed microwave sky~\cite{Bennett2003fg,Tegmark2003,Eriksen_2004,Leach2008}. Some sky regions are highly contaminated because of high foreground emission. So even after applying these cleaning algorithms, one cannot completely remove foreground contamination, entirely, from the CMB sky, particularly in the galactic plane. So, galactic masks are employed to exclude regions that may still have residual foregrounds present in the recovered CMB sky. Thus, unbiased cosmological inferences can be made using only those regions that are considered \emph{clean/cosmic} after the application of a mask.

Many deviations from the expectations of the standard cosmological model based on the cosmological principle, particularly that break isotropy, were seen in the CMB sky which have come to be known as \emph{CMB anomalies} in the full-sky CMB measurements from NASA's WMAP and ESA's \planck\ satellite missions.
These were studied by respective collaborations with almost the same level of significance~\cite{wmap7yranom,plk2013isostat,plk2015isostat,plk2018isostat}. See, for example, Refs.~\cite{schwarz2016,bull2016,abdalla2022,aluri2023cp} for a review of these CMB anomalies and other tensions in the current standard model of cosmology. All of these anomalies were found to be present even after a galactic mask was used. Some of them were also found to be present with some significance in the reasonably well measured $E$-mode polarization of the CMB sky also~\cite{Shi2022}. These indicate a violation of isotropy on large angular scales of the CMB sky, suggesting a preferred direction for our universe.

In this work, we study the unmasked regions of the CMB sky to test whether there are any unknown foreground residuals still present which could be the source of some of these discrepancies with the standard model expectations. The possible influence of residual foreground contamination on various CMB anomalies was studied earlier, for example, in Refs.~\cite{Bielewicz05,Cruz2006,Chiang2007,Bunn2008,Copi2009,Aluri2011,Hansen2012}. Here, we undertake a real (pixel) space cross-correlation analysis of the cleaned CMB sky with some known foreground templates to assess how clean the CMB map is outside the galactic mask.

The remainder of the paper is organized as follows. In Sec.~\ref{sec:ccc-stat}, we present the cross-correlation coefficient (CCC) statistic used to probe the CMB sky locally for any spurious foreground contamination. In Sec.~\ref{sec:data-sim}, the data and complementary simulations used in the present study are discussed. Then in Sec.~\ref{sec:anlys-res}, our results are presented. Finally, we conclude in Sec.~\ref{sec:concls}.

\section{CCC Statistic}
\label{sec:ccc-stat}
Here, we use a cross-correlation coefficient (CCC) statistic~\cite{Aluri2016} to locally probe regions that may still be contaminated by astrophysical emission. Various tracers of foreground emission, viz., foreground templates, are correlated with cleaned CMB sky from observations outside the masked region where the CMB signal is deemed cosmic. We do so by defining a circular disc of radius `$r$' at different locations on the sky and compute the CCC locally between a cleaned CMB sky and a foreground tracer/template. In this way we make a \emph{map} of CCC by defining circular discs covering the entire sky to check for any spurious correlations by angular scale. A patch of any shape can be used. However, for simplicity, we use a circular disc of some chosen radius.

\healpix\footnote{\url{https://healpix.sourceforge.io/}} pixelization scheme provides a framework for representing data on a sphere. A high-resolution map with more information is represented by the \healpix's map resolution parameter \nside\ where the entire sky is partitioned into 12 regions with each region subdivided into \nsidesq\ pixels. Thus, the number of pixels, \npix, in a discretized sky map is related to the \nside\ parameter as \npix=$12\times$\nsidesq. So, a higher \nside\ map has more number of pixels that can store more information on small angular scales. Once a sky signal is digitized in his way, we can use the pixel index say `$p$' after discretization and the direction $\hat{n}=(\theta,\phi)$ of an incoming CMB photon (or from any other source in the sky) interchangeably.

Thus we probe the presence of anomalous foreground residuals in a clean CMB map using CCC statistic that is defined as,
\begin{equation}
    \mathcal{R}_r(P) = \frac{ \sum_{p\in r} \left[T(p) - \bar{T}_r\right]\left[F(p) - \bar{F}_r\right]}
    {\sqrt{ \sum_{p\in r} \left[T(p) - \bar{T}_r\right]^2 \sum_{p\in r} \left[F(p) - \bar{F}_r\right]^2}}\,,
	\label{eq:ccc}
\end{equation}
where `$P$' corresponds to pixel index of CCC map whose pixel center is used to define a circular disc of radius `$r$', and `$p$' corresponds to pixels of a CMB sky ($T$) or the foreground template ($F$) that are being correlated locally using pixel values of respective maps falling within a circular patch (disc) of radius `$r$'. $\bar{T}_r$ and $\bar{F}_r$ are the mean of pixel values of cleaned CMB map and foreground templates, respectively, from that circular disc region of size `$r$'. Thus the CCC map will have values ranging from $-1$ to $+1$ as per its definition.

In applying this local CCC method, there are some practical considerations in implementing it. The circular disc defined at some sky location/pixel index `$P$' is taken in conjugation with the galactic mask. We do so to exclude the contaminated pixels as defined by the galactic map in the calculation of the CCC statistic. Once the circular disc patch of radius `$r$' is multiplied by the galactic mask, there will be regions that are excised by the galactic mask, and thus fewer pixels will be available for our analysis, especially near the galactic plane. We discard regions that do not meet the minimum pixel fraction criterion of 80\% from further analysis to keep the local CCC computation robust. Also, in order to have better estimates of the mean of CMB sky and foreground templates, we compute them from the entire sky after masking.

As explained in the next section, we generate the CCC maps at \nside=256 from a cleaned CMB map and various foreground templates that are also generated at \nside=256. Of course, we will not be able to compute CCC values at all locations/pixels due to galactic mask and partial masking of the circular disc that lead to not satisfying the pixel fraction criterion described above. Such locations / pixel values in CCC map are set to be the \healpix\ missing pixel or bad value ($-1.6375 \times 10^{30}$). A similar cross-correlation technique was used earlier in Ref.~\cite{Verkhodanov2009}.

\section{Data Sets and Simulations}
\label{sec:data-sim}
\subsection{Observational data}
In this study, we use a cleaned CMB map from \planck's 2018 data release (public release 3/PR3), cleaned using the \smica\ procedure~\cite{smica2003,smica2008,plk2018compsep} that is routinely used in many analysis. \smica\ CMB map from \planck's PR3 is available at \healpix\ \nside=2048 with a Gaussian beam of full width at half-maximum, FWHM=$5'$ (arcmin). As described later, the foreground templates are available at different beam and pixel resolutions. We process all the maps including the cleaned CMB map to have the same resolution parameters viz., \nside=256 and beam smoothing window function given  by a Gaussian beam of FWHM=$1^\circ$.
The process of downgrading the CMB map is done in harmonic space as,
\begin{equation}
    a_{lm}^{\rm out} = \frac{b_l^{\rm out} p_l^{\rm out}}{b_l^{\rm in} p_l^{\rm in}} a_{lm}^{\rm in}\,,
\label{eq:alm-con-decon}
\end{equation}
where $a_{lm}^{\rm in}=a_{lm}^{2048}$ are the spherical harmonic coefficients of the \smica\ 2018 CMB map as provided by the \planck\ team at \nside=2048 with beam transfer function $b_l^{\rm in} = b_l^{5'}$ and pixel window function $p_l^{\rm in}=p_l^{2048}$. Similarly $a_{lm}^{\rm out} = a_{lm}^{256}$ are downgraded spherical harmonics with $b_l^{\rm out} = b_l^{1^\circ}$ and $p_l^{\rm out} = p_l^{256}$. After this deconvolution and convolution process, $a_{lm}^{\rm out}$ are synthesized at \nside=256 with multipole information up to $l_{\rm max} = 2\times 256 =512$.

Now in order to omit regions that are (potentially) contaminated by foregrounds from the analysis, especially in the galactic plane, we use the \smica\ 2018 confidence mask provided along with \smica\ cleaned CMB temperature anisotropy map. Since the \smica\ CMB map and all foreground tracers are synthesized at \nside=256, the \smica\ PR3 mask is also downgraded to \nside=256 from its original resolution of \nside=2048 as follows. 
First, the \smica\ mask given at \nside=2048 is downgraded to \nside=256, and smoothed with a Gaussian beam of $1^{\circ}$ FWHM which is the same as the beam resolution of the CMB map. Then, a cut-off of 0.9 is applied on the downgraded, smoothed mask such that all the pixel values $<0.9$ are replaced with `0' and those pixel values $\geq0.9$ are replaced by `1' to get a binary map. The unmasked sky fraction of the mask thus obtained is $f_{sky} = 0.8234$ at \nside=256.

\begin{figure}
    \includegraphics[width=0.48\textwidth]{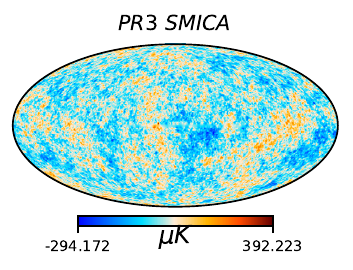}
    \includegraphics[width=0.48\textwidth]{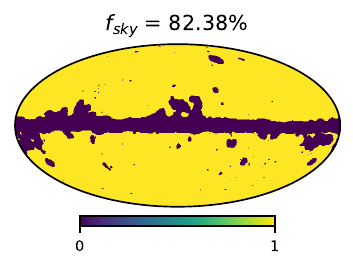}
    \caption{\emph{Left}: \smica\ 2018 CMB map at \nside=256 with Gaussian beam resolution of FWHM=$1^{\circ}$.
    \textit{Right}: \smica\ 2018 mask also at \nside=256 used in the present study.
    Both these maps are originally provided at \nside=2048 by the \planck\ collaboration.
     }
    \label{fig:cmb-mask-ns256}
\end{figure}

In Fig.~[\ref{fig:cmb-mask-ns256}], the smoothed, downgraded \smica\ CMB map from \planck\ PR3 with a Gaussian beam FWHM=$1^\circ$ and \nside=256 is shown in the \emph{left} panel, and the \smica\ mask also synthesized at \nside=256 that will be used in this work is shown in the \emph{right} panel of same figure.

For galactic foreground emission, we consider three types of components, viz., thermal dust, free-free, and synchrotron emission. We chose two templates as tracers that correspond to each of these components. Next, we briefly describe each of these galactic emission types and the templates used as their proxies. The reader may consult, for example, Refs.~\cite{MicrowaveForegrounds1999,Fauvet2012,Delabrouille2013PSM,Ichiki2014PTEP,Dickinson:2016,Angelica2008gsm1,Tegmark2017gsm2,Dunkley2017pysm,Chluba2017moment,Huang2019} for a comprehensive review on the CMB foregrounds.

\begin{figure}
\includegraphics[width=\textwidth]{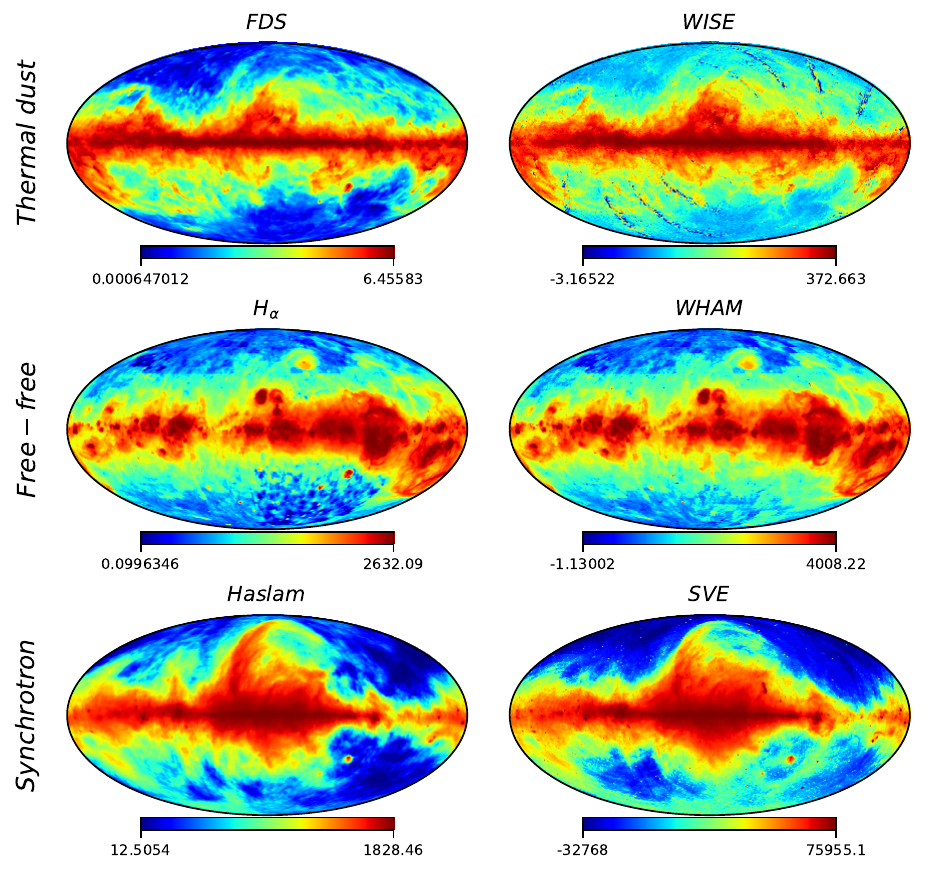}
    \caption{Templates for Galactic foreground emissions corresponding 
 to thermal dust (FDS and WISE), free-free ($H\alpha$ and WHAM), and synchrotron (Haslam and SVE) emissions in mollweide projection.
 Some relevant template properties are listed in Table~\ref{tab:temp_table}.}
    \label{fig:templates}
\end{figure}

The synchrotron emission primarily arises from cosmic-ray electrons spiraling in the magnetic field of the Milky Way. This emission dominates the sky at low microwave frequencies (below 30 GHz). We take the CMB observations at around tens or hundreds of GHz, where the synchrotron spectral energy distribution (SED) rapidly decreases with increasing frequency. Detailed models and observations indicate that at frequencies higher than a few GHz, a power-law can closely approximate the spectrum of synchrotron emission. In this case, we have used 408~MHz Haslam\footnote{\url{https://lambda.gsfc.nasa.gov/product/foreground/fg_2014_haslam_408_info.html}} map dominated by synchrotron emission from our Galaxy~\cite{Haslam1981,Haslam1982,Remazeilles2015}, and Stockert and Villa-Elisa (SVE\footnote{\url{https://lambda.gsfc.nasa.gov/product/foreground/fg_stockert_villa_info.html}}) 1.4~GHz Continuum Map as templates~\cite{W.Reich1982,P.Reich1986,Testori2001,P.Reich2001}.

Thermal dust emission dominates the foreground contribution at frequencies above 70~GHz. We know that apart from Hydrogen and electrons, the interstellar medium (ISM) is also filled with dust grains mostly made of graphites, silicates, and Polycyclic Aromatic Hydrocarbons (PAHs), whose size (diameter) ranges from less than a nanometer to roughly a micron. The exact grain composition may vary with the local environment. When these tiny dust grains are heated by interstellar radiation, they thermally re-emit this energy, which is called thermal dust emission. We use Finkbeiner, Draine \& Schlegel (FDS) model dust template\footnote{\url{https://lambda.gsfc.nasa.gov/product/foreground/fg_fds_info.html}} at 94~GHz~\cite{Schlegel_1998,Finkbeiner_1999}, and Wide-field Infrared Survey Explorer (WISE\footnote{\url{https://lambda.gsfc.nasa.gov/product/foreground/fg_wise12_micron_dust_map_info.html}})~\cite{Meisner_2013} 12-micron sky map as templates for diffuse thermal dust emission from our galaxy.

\begin{table}
	\centering
	\begin{tabular}{lccc} % four columns, alignment for each
		\hline
		Templates & \nside\ & Resolution(arcmin) & Unit\\
		\hline
		FDS & 512 & 6.1 & mK\\
		WISE & 1024 & 12 & MJy/sr\\
		  $H\alpha$ & 1024 & 6 & Rayleighs\\
            WHAM & 256 & 60 & Rayleighs\\
            Haslam & 512 & 56 & K\\
            SVE & 256 & 35.4 & mKTB\\
		\hline
	\end{tabular}
	\caption{Some characteristics of Galactic foreground templates that we used in this work.}
	\label{tab:temp_table}
\end{table}

Free-free emission (or thermal bremsstrahlung) arises due to scattering between free electrons and ionized Hydrogen (protons) in the interstellar plasma.
This emission can be traced with the $H\alpha$ line emission, which is predominantly emitted from regions of ionized hydrogen (HII) or active star-forming regions in the galaxy. For CMB experiments, free-free emission is particularly significant because it is the only foreground component that remains non-negligible across all frequencies between 1 and 100~GHz, making it especially sensitive to degeneracies with other foreground components viz., synchrotron.
Unlike thermal dust and synchrotron radiation, free-free is not dominant at any of the frequencies where it can be measured to use as a template. Therefore, $H\alpha$ maps are used as tracers of free-free emission~\cite{Valls-Gabaud1998}.
In this work, we use Composite all-sky  $H\alpha$ map\footnote{\url{https://lambda.gsfc.nasa.gov/product/foreground/fg\_halpha\_info.html}}~\cite{Finkbeiner2003}, and the WHAM (Wisconsin H-alpha Mapper) map\footnote{\url{https://lambda.gsfc.nasa.gov/product/foreground/fg_wham_h_alpha_map_info.html}}~\cite{Haffner2003whamnorth,Haffner2010whamsouth} as proxies for free-free emission.

\begin{figure}
\centering
\includegraphics[width=0.85\textwidth]{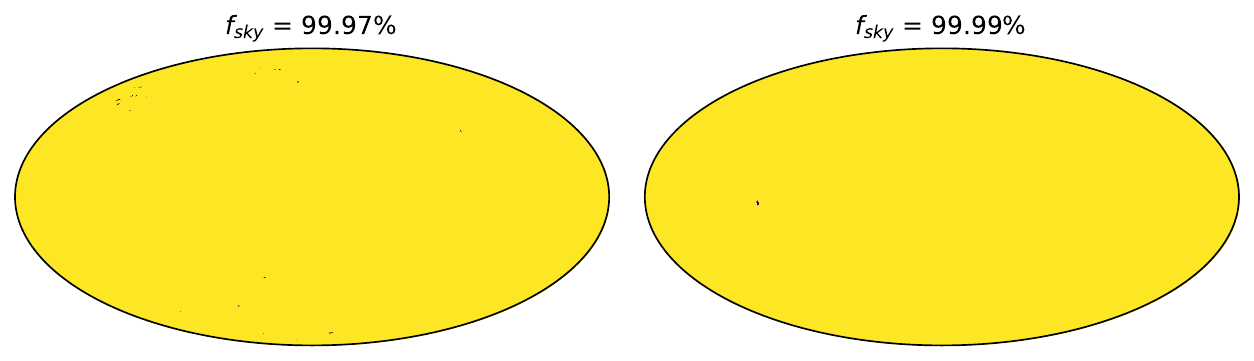}
\includegraphics[width=0.85\textwidth]{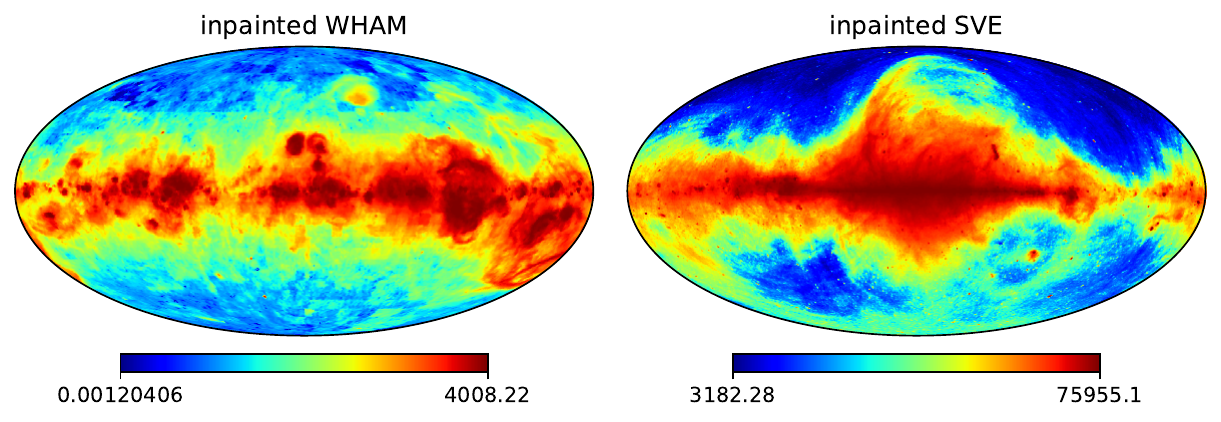}
\includegraphics[width=0.85\textwidth]{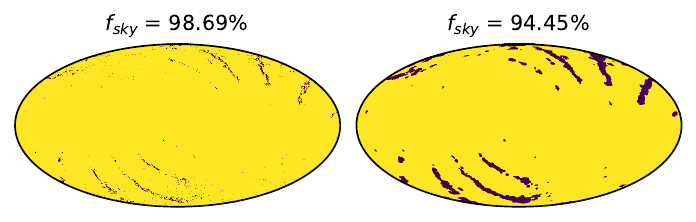}
    \caption{\emph{Top} : Mask corresponding to missing or negative pixels in WHAM and SVE foreground templates that are filled using diffuse inpainting method~\cite{bucher2012}.
    \emph{Middle} : WHAM and SVE templates after inpainting.
    \emph{Bottom} :
    Missing or negative pixels in WISE dust template are shown in the \emph{bottom left} panel. They form stripe-like features as also evident from \emph{top right} panel of Fig.~[\ref{fig:templates}].
    An extended WISE mask is shown in the \emph{bottom right} panel to handle the missing data and map artifacts. The process of extending the \emph{bottom left} mask is explained in the text.}
   \label{fig:mask_wham_sve_wise}
\end{figure}

Hereafter, we refer to the two synchrotron templates as Haslam and SVE, the two thermal dust templates as FDS and WISE, and the free-free emission templates as Composite $H\alpha$ and WHAM maps. Some of the properties of these templates/maps, viz. \nside, (effective Gaussian) beam resolution and map units, are listed in Table~\ref{tab:temp_table}, and are shown in Fig.~[\ref{fig:templates}]. The color-bar of WISE, WHAM, and SVE shows that these templates have some missing pixel/negative values. In case of WHAM and SVE, we masked such pixels and then filled them using the diffuse inpainting method~\cite{bucher2012}.
The masks generated by thresholding the WMAH and SVE maps to eliminate such negative or missing pixels for diffuse inpainting are shown in \emph{top row} of Fig.~[\ref{fig:mask_wham_sve_wise}]. As is evident, there are very few stray pixels that need to be filled in. The inpainted WHAM and SVE templates after inpainting are shown in \emph{middle row} of the same figure.

On the other hand, the WISE foreground template has some arc-like stripes, as evident from \emph{top right} panel of Fig.~[\ref{fig:templates}]. In this case, we were unable to suitably eliminate the stripes and inpaint them. So, we use masked WISE map in our analysis at \nside=1024 as provided. First to identify the regions where there are missing pixel or pixels with negative values, we applied a cut-off on the WISE map so that those that are $\leq0.01$ are set to `0' and `1' otherwise. The binary map thus obtained is shown in \emph{bottom left} panel of Fig.~[\ref{fig:mask_wham_sve_wise}]. We can see that missing pixels form stripe-like structures in the sky. (We note that similar artifacts are present in the WISE map apart from these which we tried to eliminate.) Then we inverted this binary mask, downgraded it from \nside=1024 to \nside=128 and applied a Gaussian beam of FWHM=$80'$, and thresholded this smoothed mask such that those pixels whose pixel value is $\leq 0.1$ is set to `0' and `1' otherwise. Now we upgraded this thresholded mask from \nside=128 to \nside=1024 and applied a Gaussian beam of $60'$ FWHM to smooth out edge effects upon upgradation, and changed the values of those pixels that are $\leq 0.1$ to `0' and `1' otherwise. Finally, this upgraded mask is inverted that will be combined with \smica\ mask for use in our analysis. This extended mask to be used with WISE template is shown in the \emph{bottom right} panel of Fig.~[\ref{fig:mask_wham_sve_wise}].

The original \nside\ and beam FWHM of the individual templates are different. Therefore, we downgraded all templates, except WISE map, to a common \healpix\ pixel resolution of \nside=256 and beam resolution of $1^{\circ}$ FWHM following Eq.~(\ref{eq:alm-con-decon}). As mentioned above, because we are unable to fill the survey artifacts in WISE template, we use it at its native resolution of \nside=1024 and FWHM=$10'$. Similarly, the \smica\ cleaned CMB map from PR3 and corresponding simulations were synthesized at \nside=256 with beam FWHM=$1^{\circ}$, except when correlating with the WISE template. For this specific template we generated all maps - data and simulations - at \nside=1024 and a map smoothing level given by a Gaussian beam of FWHM=$10'$. However, the CCC maps and the corresponding $p$-value maps are all derived at \nside=256.

These templates are used as proxies for respective diffuse galactic emission types that are created from observations taken at frequencies where they are dominant. As already mentioned, synchrotron is the dominant source of microwave emission from our galaxy at low frequencies ($\lesssim 30$~GHz), whereas thermal dust is dominant at high frequencies ($\gtrsim100$~GHz). However, free-free emission is not dominant by itself in the microwave frequency regime where WMAP and \planck\ made observations. So, $H\alpha$ emission that is also emitted from the same HII regions is taken as a tracer for free-free emission, whose intensity ($I_{H\alpha}$) is related to free-free emission. These foregrounds measured at some frequency ($\nu_0$) where they are dominant are extrapolated to other frequencies by a power law as $T_A (\hat{n}) = F_{\nu_0} (\hat{n}) (\nu/\nu_0)^\beta$, where $F_{\nu_0} (\hat{n})$ is the observed foreground sky signal at some reference frequency $\nu_0$ and $T_A$ is the antenna temperature in which foregrounds are usually modeled\footnote{They can be converted to thermodynamic units by a multiplicative factor. The WMAP and \planck\ collaborations gave these conversion factors for the frequency bands in which they made observations~\cite{Bennett2003fg,Bennett2013finalmaps,Planck:2013lfi,Planck:2013hfi,Planck:2015fg}}~\cite{BouchetGispert1999}. For synchrotron the spectral index is steep with $\beta_{s,\rm all-sky}\sim -3.1$ where as for free-free its taken to be $\beta_f=-2.14$. Depending on a particular region in the sky, synchrotron and free-free cannot be disentangled due to closer spectral index values. Thermal dust is modeled with a spectral index of $\beta_d\sim 1.8$. However, since HFI bands of \planck\ go beyond the WMAP frequency coverage, they are modeled across \planck\ frequencies as a gray body instead~\cite{Planck:2015fg}.

\subsection{Simulations}
Simulated CMB maps corresponding to the \planck\ \smica\ 2018 cleaned CMB map are provided by \planck\ Collaboration as part of their public data release 3 (PR3). Also provided are simulated noise maps to add with these CMB realizations. They are referred to as Full Focal Plane (FFP) simulations, with the third set referred to as FFP10. These simulations reflect key aspects of the \planck\ scanning strategy, detector responses, and other observational artifacts of the telescope, as well as the data reduction pipeline throughout the full-mission period. We use the FFP10 \smica\ CMB simulations added with noise from PR3. Although there are 1000 simulated CMB maps, only 300 noise realizations are provided. Hence these 300 noise maps are added to 1000 CMB maps by permuting them. Further, one of the 1000 CMB maps got corrupted (simulation no. 970). Therefore, there are only 999 CMB realizations available. Thus we generate a mock noisy CMB map by randomly selecting 10 maps (from rest of the 999 maps) from which $a_{lm}$'s are extracted from 10 non-overlapping multipole bins from each map to synthesize the 1000th simulation. More details about these simulations can be found in Ref.~\cite{plk2018isostat}.

\planck\ FFP10 simulations are provided at \nside=2048 with a Gaussian beam resolution of FWHM=$5'$. So, these are downgraded to the same pixel and beam resolution as the data i.e., they are downgraded to \healpix\ \nside=256 with a map smoothing level given by a Gaussian beam of FWHM=$1^\circ$ following Eq.~(\ref{eq:alm-con-decon}). A second set of simulations were generated to use with the WISE dust template at \nside=1024 with a resolution of FWHM=$10'$ Gaussian beam. The extended mask generated to deal with artifacts in the WISE map is shown in the \emph{bottom right} panel of Fig.~[\ref{fig:mask_wham_sve_wise}], which will be multiplied by the \smica\ 2018 PR3 mask when performing the correlation analysis with it.

\begin{figure}[t]
\centering
\includegraphics[width=0.95\textwidth]{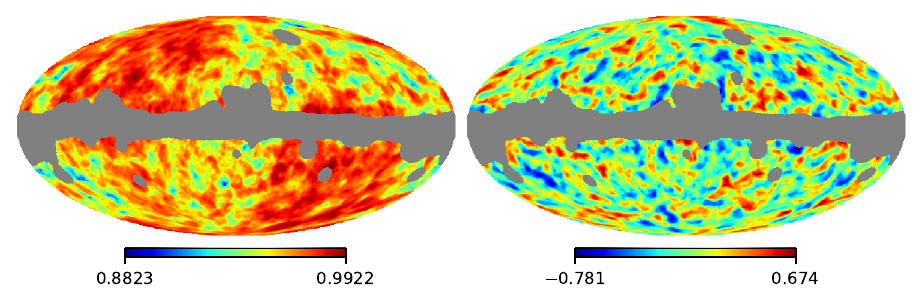}
~
\includegraphics[width=0.9\textwidth]{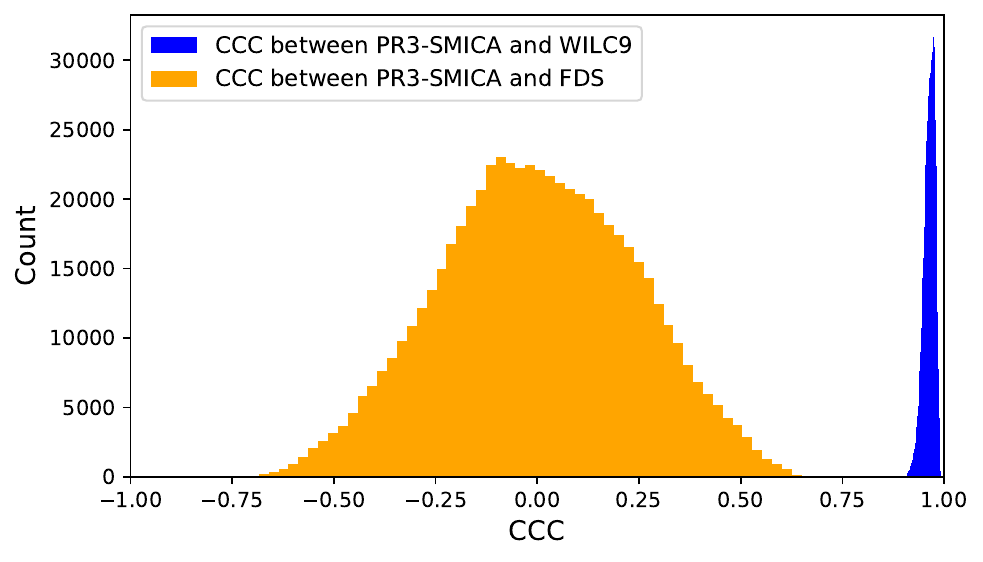}
    \caption{\emph{Top left} : CCC map obtained by correlating \smica\ CMB 2018 map (PR3-\smica) and WMAP's 9yr ILC cleaned CMB map (WILC9) for $r=5^\circ$. \emph{Top right} : CCC map from PR3-\smica\ CMB map and  FDS thermal dust cross-correlation for the same choice of disc radius. \emph{Bottom}: Histogram of the two CCC maps shown above. Their histograms visually match the expectations.}
    \label{fig:ccc_test}
\end{figure}

\section{Analysis and Results}
\label{sec:anlys-res}

Here, we present results from our CCC analysis where \smica\ cleaned 2018 CMB map is cross-correlated \emph{locally} with various tracers of galactic foreground, specifically, synchrotron, free-free, and thermal dust emission from our own galaxy. Two templates are used for each foreground type in our analysis. They are named FDS and WISE for thermal dust, $H\alpha$ and WHAM for free-free, and Haslam and SVE for synchrotron emission. As mentioned above, all data and simulations are generated at \healpix\ \nside=256 having a smoothing level given by a Gaussian beam of FWHM=$1^\circ$. However, due to the nature of WISE dust template, all data and simulations are generated at its native resolution of \nside=1024 and a beam resolution of FWHM=$10'$. The CCC maps thus derived following Eq.~(\ref{eq:ccc}) are at \nside=256.

CCC maps with various foreground tracers are derived by defining circular discs of different radii to filter the CMB map for residual foregrounds by angular size. We used 13 different disc radii viz., $r=1^\circ,2^\circ,3^\circ,4^\circ,5^\circ,6^\circ,8^\circ,10^\circ,12^\circ,15^\circ,20^\circ,25^\circ$ and $30^\circ$ centered at the pixel centers of \nside=256 \healpix\ grid covering the sky uniformly. The circular discs are taken in combination with the \smica\ 2018 galactic confidence mask so that we probe for residual foregrounds in the supposedly clean regions of the recovered CMB sky. To ensure that the CCC computed locally is estimated as robustly as possible, it is computed from a circular disc of chosen radius `$r$' defined at a particular location on the sky only if at least 80\% of the pixels survive the galactic cut (i.e., circular disc mask in conjugation with the galactic mask) compared to the total number of pixels in a full circular disc defined at that location.

As a test, we first correlate the cleaned CMB map from the WMAP 9 year data release, specifically the ILC map (internal linear combination map), with the \smica\ 2018 CMB map. Since both correspond to the same CMB sky but from different missions, we expect them to be highly correlated with the cross-correlation coefficient being (close to) `1' across the sky. The resulting CCC map for a choice of $r=5^\circ$ is presented in \emph{top left} panel of Fig.~[\ref{fig:ccc_test}]. Complementing this, we presented the CCC map derived by correlating \planck's PR3 \smica\ CMB map with FDS template tracing galactic thermal dust emission in the \emph{top right} panel for the same choice of $r=5^\circ$ disc radius. We expect the CCC values of this map are centered around `0'. The histograms of both CCC maps' pixel values are shown in the \emph{bottom} panel of the same figure. Just, as expected, the CMB maps from WMAP and \planck\ are highly correlated with their CCC map's histogram clustering at `1'. And with the FDS template, the CCC map histogram shows a peak at around `0', indicating no correlation, but also regions that are positively and negatively correlated. Of course, this is a qualitative / visual assessment. They have to be quantified with simulations as presented later.

\begin{figure}
	\centering
    	\includegraphics[width=0.95\textwidth]{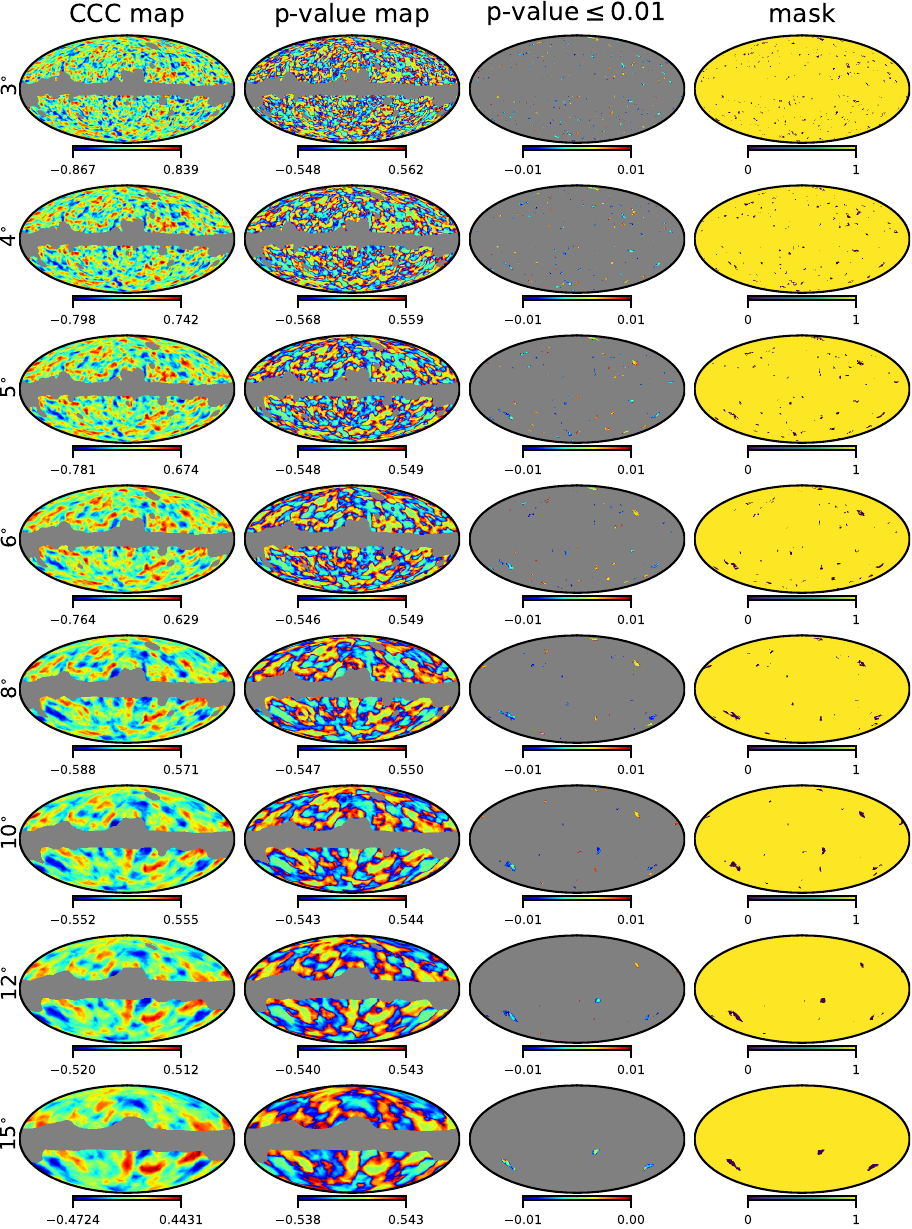}
    \caption{\emph{$1^{st}$ column}: CCC maps corresponding to PR3 \smica\ and FDS dust map for $r=3^{\circ}, 4^{\circ}, 5^{\circ}, 6^{\circ}, 8^{\circ}, 10^{\circ}, 12^{\circ}$ and $15^{\circ}$ disc radii. \emph{$2^{nd}$ column}: $p$-value maps created in comparison with 1000 simulations. Here negative sign represent the $p$-value for anti-correlation. \emph{$3^{rd}$ column}: Anomalous regions identified with $p$-value cut-off of $|p|\leq0.01$. \emph{$4^{th}$ column}: Binary masks obtained from 1\% $p$-value cut maps shown in the $3^{rd}$ column by mapping the non-zero $p$-values to `0' and the rest to `1'.}
    \label{fig:FDS}
\end{figure}

\begin{figure}
	\centering
	\includegraphics[width=\textwidth]{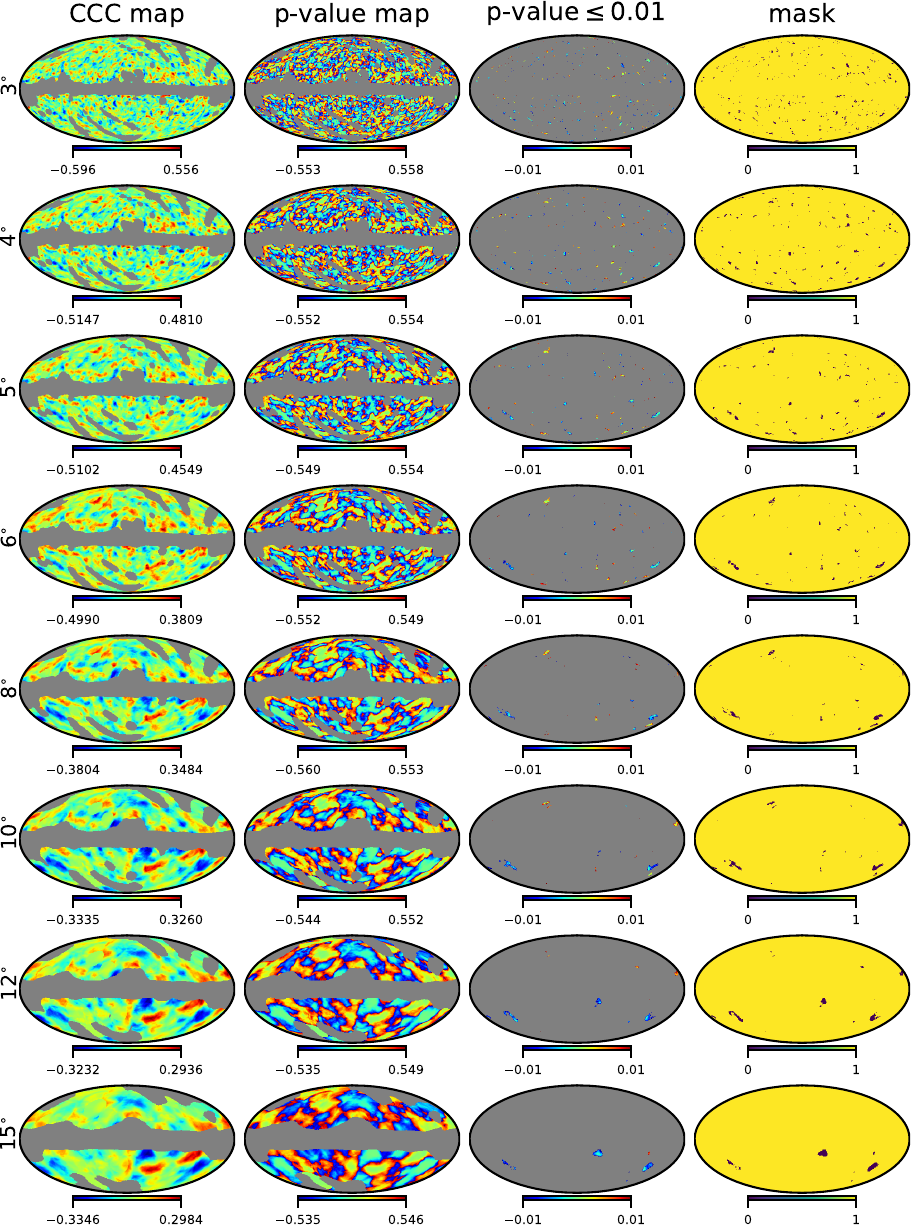}
    \caption{This figure is similar to the Fig.~\ref{fig:FDS}, but CCC analysis of PR3-\smica\ CMB map with WISE thermal dust template at \nside=1024 is presented here.}
    \label{fig:WISE}
\end{figure}

\begin{figure}
	\centering
	\includegraphics[width=\textwidth]{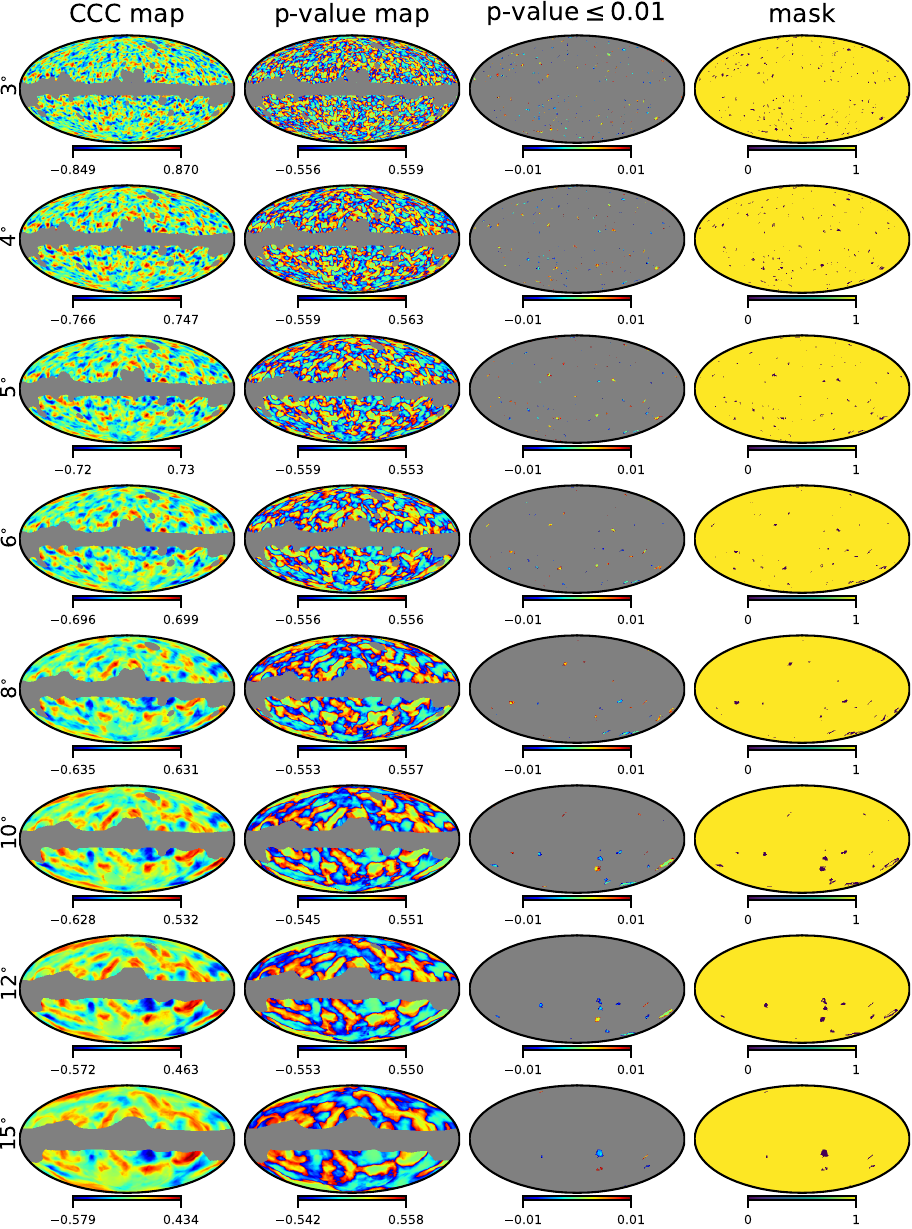}
    \caption{Same as Fig.~\ref{fig:FDS} but presented here are the results from local cross-correlation analysis of PR3-\smica\ CMB map with $H\alpha$ free-free proxy.}
    \label{fig:halpha}
\end{figure}

\begin{figure}
	\centering
	\includegraphics[width=\textwidth]{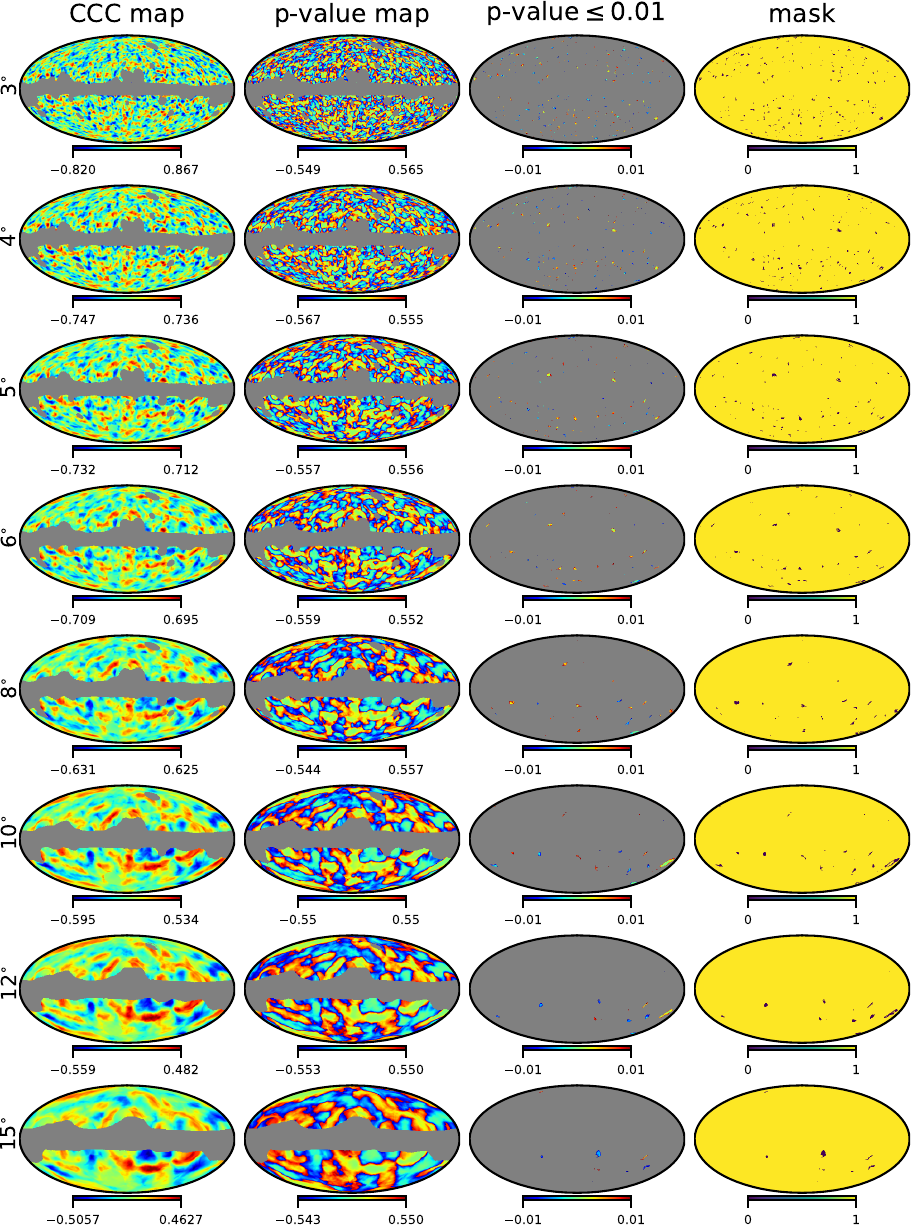}
    \caption{Same as Fig.~\ref{fig:FDS} but for PR3-\smica\ CMB map correlated with WHAM free-free template.}
    \label{fig:WHAM}
\end{figure}

\begin{figure}
	\centering
	\includegraphics[width=\textwidth]{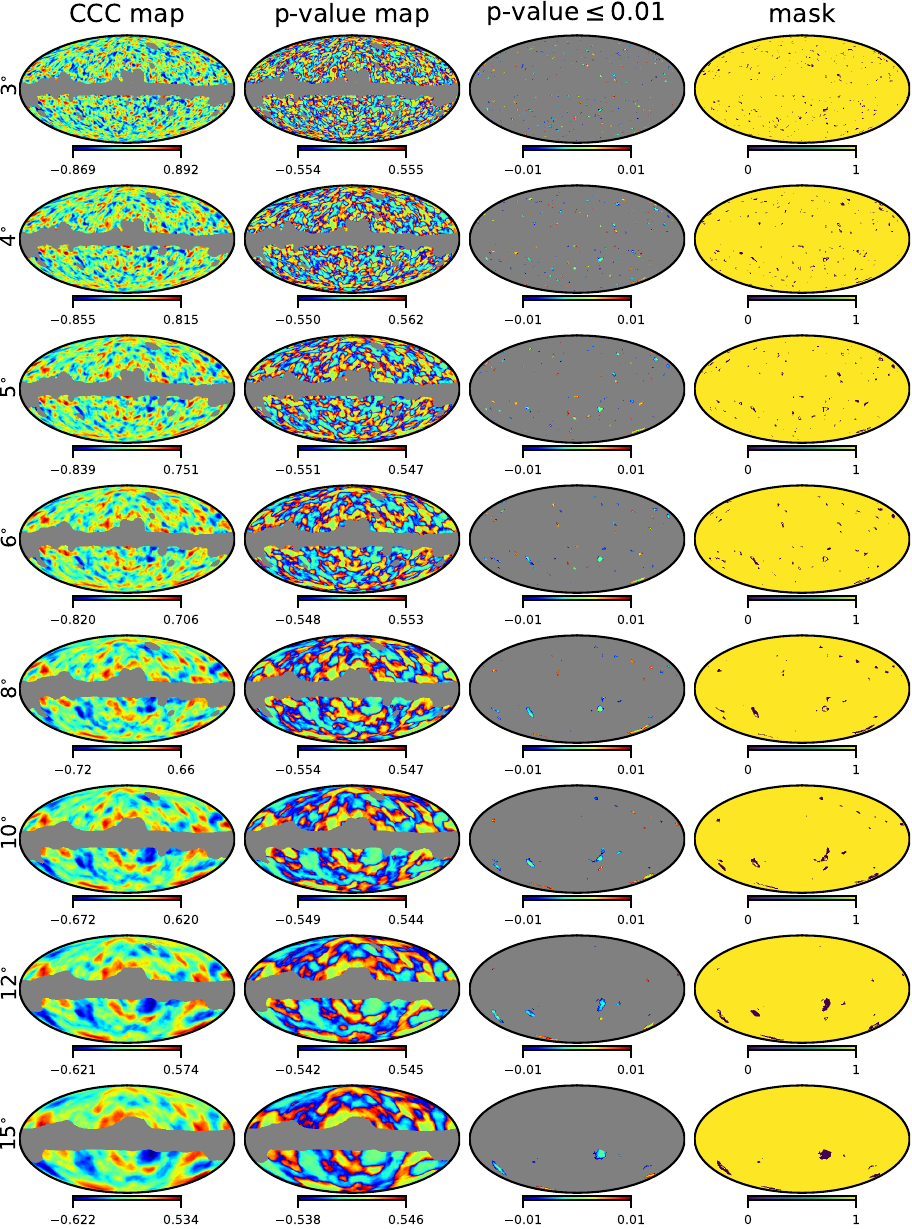}
    \caption{Same as Fig.~\ref{fig:FDS} but results from CCC analysis of PR3-\smica\ CMB map with Haslam 408~MHz map dominated by galactic synchrotron emission.}
    \label{fig:haslam}
\end{figure}

\begin{figure}
	\centering
	\includegraphics[width=\textwidth]{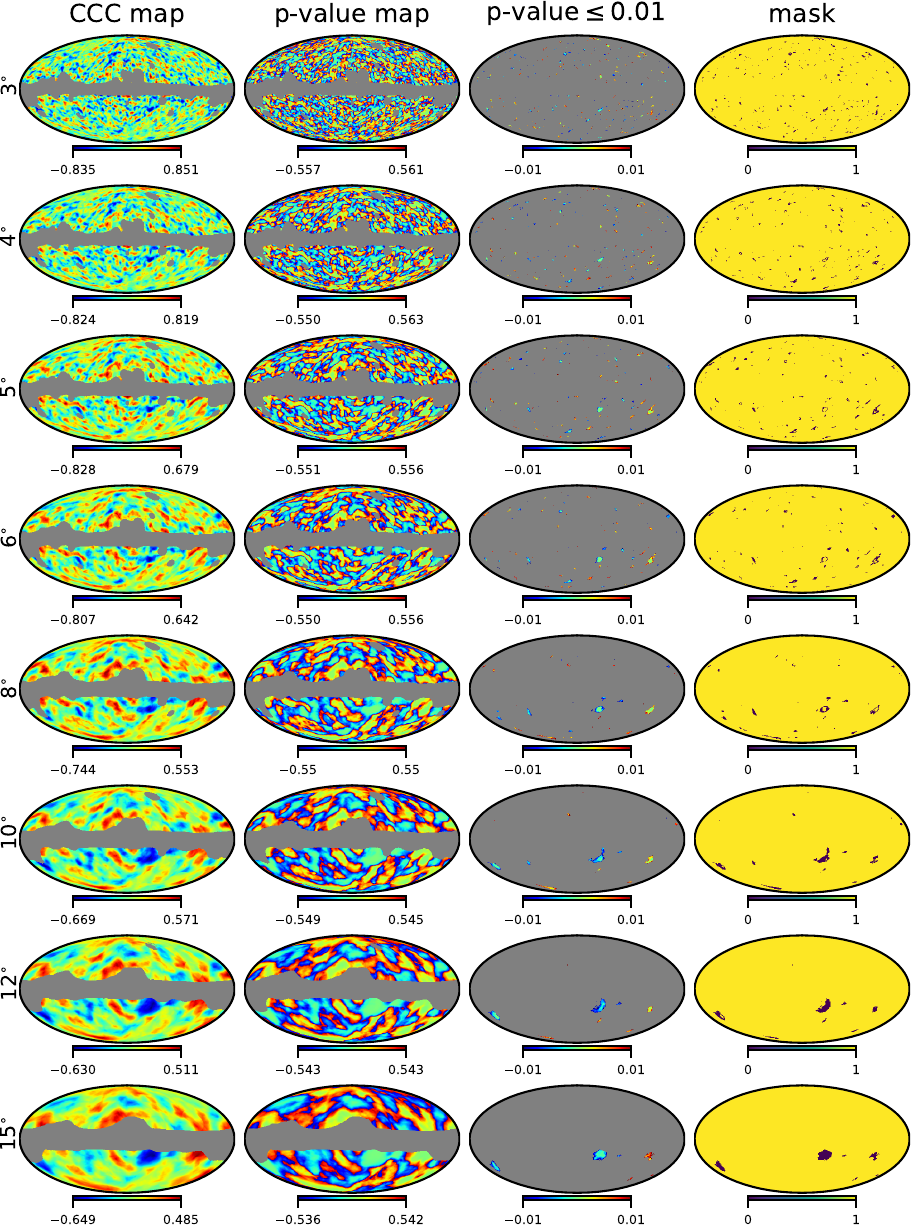}
    \caption{Same as Fig.~\ref{fig:FDS} but for PR3-\smica\ CMB map correlated locally with SVE synchrotron tracer using different disc radii to probe foreground residuals by angular size.}
    \label{fig:SVE}
\end{figure}

Now we go on to show our results from CCC analysis of correlating PR3-\smica\ CMB map with the thermal dust templates viz., FDS, WISE maps in \emph{first} column of Fig.~[{\ref{fig:FDS}}] and [\ref{fig:WISE}], with free-free templates i.e., $H\alpha$ and WHAM maps again in \emph{first} column of Fig.~[\ref{fig:halpha}] and [\ref{fig:WHAM}], and finally with galactic synchrotron tracer viz., Haslam and SVE templates also in \emph{first} column of Fig.~[\ref{fig:haslam}] and [\ref{fig:SVE}], respectively. As mentioned above, although we performed the CCC analysis for 13 different choices of disc radii, we presented results only from $r=3^\circ$ to $15^\circ$, for reasons explained later. From these figures it is clear that the CCC maps take both positive and negative values, with most of the CCC values \emph{appearing} to be (nearly) zero across the sky from the color-bar.

Such a visual inspection is not sufficient.
Similar to the data, the simulated noisy CMB maps are correlated with the same foreground templates to get quantitative inferences. (Recall that the WISE dust template is correlated with simulations generated at \nside=1024 with beam resolution given by Gaussian kernel of FWHM=$10'$, while others are generated at \nside=256 with Gaussian beam of FWHM=$1^\circ$.) These CCC maps from simulations (all derived at \nside=256) are compared with data-derived CCC maps pixel-by-pixel to get a $p$-value map for the cross-correlation coefficient computed across the sky for different choices of disc radii with all foreground tracers. If a data CCC map's value is positive, we compute the $p$-value as the number of times the simulation CCC maps exceed that of the data CCC value. If the data CCC value is negative, we count the number of times the simulated CCC maps were found to have more negative CCC values compared to that in data in that pixel direction to get $p$-value maps. They are shown in the \emph{second} column of Fig.~[\ref{fig:FDS}], [\ref{fig:WISE}], [\ref{fig:halpha}], [\ref{fig:WHAM}], [\ref{fig:haslam}] and [\ref{fig:SVE}] respectively. Note that we assigned a negative $p$-value to regions that are anticorrelated and a positive $p$-value to positively correlated region/pixel in data CCC maps.

\begin{figure}
    \includegraphics[width=\columnwidth]{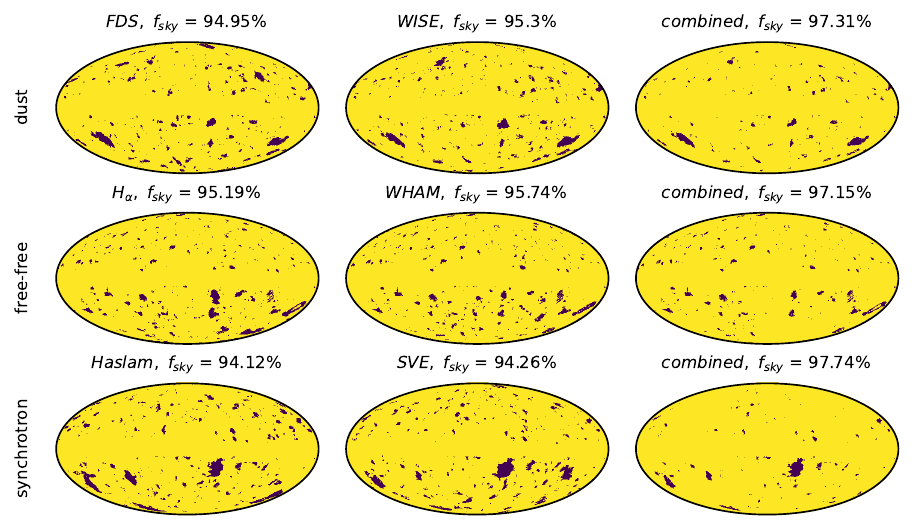}
    \caption{
    Masks shown in \emph{first} and \emph{second} column are obtained by combining all the binary masks depicted in the $4^{th}$ column of Fig.~[\ref{fig:FDS}] (FDS), Fig.~[\ref{fig:WISE}] (WISE), Fig.~[\ref{fig:halpha}] ($H\alpha$), Fig.~[\ref{fig:WHAM}] (WHAM), Fig.~[\ref{fig:haslam}] (Haslam), and Fig.~[\ref{fig:SVE}] (SVE) for $r=3^\circ$ to $15^\circ$ disc radii.
    In the \emph{third} column, the combined individual foreground masks from the two tracers of a specific emission type as mentioned are shown. They ($3^{rd}$ column) are obtained by performing an \texttt{.and.} operation on the masks shown to the left i.e., $\{0,0\}\rightarrow0$ and $\{0,1\}$ or $\{1,0\}$ or $\{1,1\} \rightarrow 1$.} 
    \label{fig:cmask}
\end{figure}

A wide range of $p$-values were found for the CCC maps from data when compared with simulations. In order to identify regions that have spurious foreground residuals in the cleaned CMB sky from observations, we impose a $p$-value cut-off of 1\% or less to define anomalously correlated regions. The regions in the $p$-value maps for each foreground tracer that have a $p$-value of $\leq0.01$ (positive or negative) are separately shown in the \emph{third} column of respective figures. In this way, we get eight masks for every foreground template from disc radius $r=3^\circ$ to $15^\circ$. As we can see, the $p$-value maps with $p$-value $\leq 0.01$ are mostly empty. But still, there are some small regions that show significant contamination by foreground residuals in the cleaned CMB map. For smaller disc radii, we find that there are many anomalously correlated small island-like regions or stray pixels. But, for larger discs radii, there are few blob-like anomalous regions. Further, we also see some anomalous regions that are common in the $p$-value maps with $p$-value $\leq 0.01$ in Fig.~\ref{fig:FDS} to Fig.~\ref{fig:SVE}. 

Now, those regions that are deemed anomalous by the 1\% $p$-value cut are shown as binary masks by angular size in the \emph{fourth} column of Fig.~[\ref{fig:FDS}] to Fig.~[\ref{fig:SVE}] for various tracers in the same order as previously explained. Here we note that the $p$-value maps with a $p$-value cut-off of 1\% or the binary maps derived thereof show a lot of stray pixels for disc radii $r=1^\circ$ and $2^\circ$. These disc radii are close to the smoothing level of the input data maps (and simulations) and could potentially be random noise with no information at those scales. So, we present our results starting from $r=3^\circ$ (=$3\times$FWHM=$3\times1^\circ$) in Fig.~[\ref{fig:FDS}] to Fig.~[\ref{fig:SVE}]. On the other hand, for $r>15^\circ$, we see no extended regions that are spuriously correlated with foregrounds (as per the 1\% $p$-value cut we chose). So we didn't present CCC maps or corresponding $p$-values beyond $r=15^\circ$ disc radius. It is also expected that such large swaths of sky will not have unknown foreground contamination still present in the cleaned CMB sky.

Since the $p$-value maps from each of the two tracers for a particular galactic foreground correspond to the same emission type viz., thermal dust in Fig.~[\ref{fig:FDS}] and [\ref{fig:WISE}], free-free in Fig.~[\ref{fig:halpha}] and [\ref{fig:WHAM}] and synchrotron in Fig.~[\ref{fig:haslam}] and [\ref{fig:SVE}], they look similar but not identical. The effective binary masks for each tracer combining (multiplying) the individual binary masks for $r=3^\circ$ to $15^\circ$ shown in the \emph{fourth} column of Fig.~[\ref{fig:FDS}] to [\ref{fig:SVE}] are exhibited in the \emph{first} and \emph{second} columns of Fig.~[\ref{fig:cmask}]. 
Thus we create a single mask for each foreground if a CCC value in the CCC map from both tracers of an emission type is simultaneously anomalous (per our 1\% $p$-value cut criterion). They are shown in the \emph{third} column of Fig.~[\ref{fig:cmask}].

\begin{figure}
\centering
\includegraphics[width=0.45\textwidth]{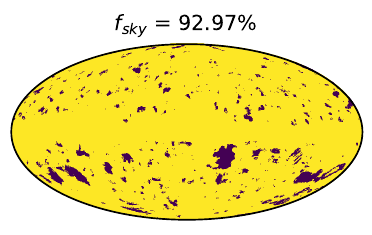}
~
\includegraphics[width=0.45\textwidth]{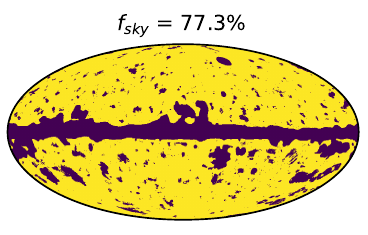}
    \caption{\emph{Left}: Newly created mask to exclude regions outside the galactic mask with spurious foreground correlations, that we call SP-FG-CORR mask, after combining the three masks of $3^{rd}$ column of Fig.~[\ref{fig:cmask}]. \emph{Right}: SP-FG-CORR mask combined with \planck\ 2018 \smica\ CMB mask is shown. Respective sky fractions are also indicated.}
    \label{fig:combMask}
\end{figure}

Finally, a common mask is obtained by combining all the three individual \emph{foreground} masks of Fig.~[\ref{fig:cmask}] (third column). That combined mask excising regions with spurious foreground correlations (say SP-FG-CORR mask) outside the galactic mask is shown in the \emph{left} panel of Fig.~[\ref{fig:combMask}]. Through this CCC analysis we find that about 7\% of the sky may be still contaminated in the supposedly clean regions of CMB sky. We advocate that this mask should be used in conjugation with the \planck\ provided galactic masks to avoid potential biases due to use of information from regions that are strongly correlated with foregrounds in any cosmological analysis. The same SP-FG-CORR mask combined with \smica\ 2018 galacic mask is shown in the \emph{right} panel of the Fig.~[\ref{fig:combMask}]. This proposed new mask has an available sky fraction of $f_{sky}\approx77.3\%$. Note that \smica\ 2018 galactic mask shown in \emph{right} panel of Fig.~[\ref{fig:cmb-mask-ns256}] has a non-zero sky fraction of $f_{sky}\approx82.38\%$. So an additional $\sim5\%$ of the sky has to be omitted due to strong indications of residual foreground contamination from our analysis.

\begin{figure}
\centering
\includegraphics[width=0.75\textwidth]{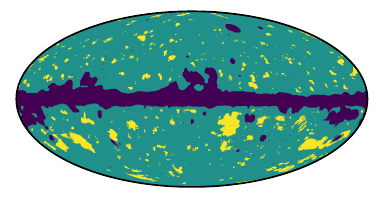}
    \caption{
    Newly excised regions that are found to be anomalously correlated with various foreground tracers are highlighted in \emph{yellow} on the original PR3 \smica\ mask. \emph{Dark blue} regions are masked out in the original mask and those in \emph{yellow} regions are newly found regions that may be still potentially contaminated by foreground found in this study.} 
    \label{fig:newMask}
\end{figure}

\section{Conclusions}
\label{sec:concls}
In this study, our aim was to explore the potential presence of residual foregrounds in a cleaned cosmic microwave background (CMB) map. It is customary to use a galactic mask to omit regions of the recovered CMB that may still have significant foreground contamination. Here we intend to probe for foreground residuals outside a galactic mask where the estimated CMB signal is deemed clean. To accomplish this objective, we used a cross-correlation coefficient (CCC) statistic on \planck's 2018 \smica\ cleaned CMB map. It is correlated locally with foreground templates corresponding to three major galactic emissions viz., thermal dust (FDS and WISE maps), free-free ($H\alpha$ and WHAM maps) and synchrotron (Haslam and SVE maps) by defining circular discs of varying size from $r=1^\circ$ to $30^\circ$ uniformly covering the entire sky. This approach enabled the identification and filtration of anomalous regions by way of spurious correlations with the foreground templates by their angular size. These local circular disc masks were combined with the temperature mask prescribed for use with \planck\ 2018 \smica\ CMB map. Further, we imposed a pixel fraction cut where we demand at least 80 percent non-zero pixels in the surviving circular disc mask when taken in combination with the foreground mask, compared to the full circular disc mask of a selected radius.

To complement the CCC maps derived from observations, a set of 1000 simulations referred to as FFP10 simulations were used that were provided by \planck\ team as part of their public release 3 (PR3). However there are 1000 CMB and 300 noise maps that are processed according to the \smica\ cleaning procedure. Further one of the CMB realizations was corrupt. As such only 999 CMB simulations were available. We permuted the noise maps to generate 999 noisy \smica\ CMB realizations. Then, we randomly selected 10 noisy maps and combined the spherical harmonic coefficients ($a_{lm}$'s) from 10 non-overlapping multipole bins from each of these 10 maps to synthesize one more map making a simulation ensemble of 1000 (noisy) CMB maps that have all observational artifacts incorporated. CCC maps from simulations are derived following the same procedure used in deriving the data CCC maps. Thus, a total of 1000 CCC maps were obtained by locally cross-correlating the FFP10 simulations with the six foreground templates.

We find that the data CCC values when correlated with various foregrounds have both positively and negatively correlated regions along with uncorrelated regions.

Corresponding to every CCC map between CMB sky and foreground templates, to quantify the significance of a CCC value in these maps, p-values for each valid pixel at any location of the sky were computed by comparing the data statistic with the same from simulaitons pixel-by-pixel. It is done by counting the number of times a CCC map's pixel value from simulaitons exceeds (positively or negatively) compared to the data statistic. 
Then, regions with significant foreground contamination are inferred by imposing $p\leq0.01$ as a cut-off on the $p$-value maps.

By retaining only those pixels of $p$-value map that have pixel values $\leq 0.01$, we create binary masks by combining such threshodled $p$-maps from $r=3^\circ$ to $15^\circ$. Since the input maps' smoothing level is given by a Gaussian beam of FWHM=$1^\circ$, we considered combining $p$-value map after 1\% cut from $r=3^\circ$ onwards (disc radii that are three times map's smoothing level). For disc radii choices of $r\geq15^\circ$, no anomalous regions were found as expected. Binary masks from the two foreground tracers of a particular foreground emission type viz., FDS and WISE for thermal dust, $H\alpha$ and WHAM for free-free, and Haslam and SVE for synchrotron, are combined only if the CCC map's pixels are anomalous in both tracers with respect to the 1\% $p$-value cut. The three resulting masks for each emission type are combined to get a final mask that highlights regions of anomalous correlation with foreground templates outside the galactic mask. The new regions to be excluded are shown in \emph{yellow} in Fig.~[\ref{fig:newMask}], where the \emph{dark blue} region is excluded by the PR3 \smica\ mask as provided.

This new mask may be used for cosmological analysis to avoid being biased by using CMB sky from the regions highlighted in \emph{yellow} in Fig.~[\ref{fig:newMask}] that are potentially still contaminated, as revealed by our CCC analysis. The broad yellow blob to the right of galactic center below the galactic plane seems to match with one of the regions depicted in Fig.~13 of Ref.~\cite{wmap7yranom} which upon omitting indicated that the quadrupole-octopole alignment in ILC cleaned CMB map from WMAP 7yr data has changed from being (almost)perfectly aligned (top panel) to being aligned at about $12.6^\circ$ (bottom right panel) with each other. So this new mask could perhaps be relevant in performing tests of isotropy.

\section*{Acknowledgments}
PKS acknowledges financial support from Gangadhar Meher University, Sambalpur, Odisha, India that is funded by the State government of Odisha, India through the SEED research grant No.4934/GMU. In this work, we extensively used the publicly available \healpix/\healpy\footnote{\url{https://healpy.readthedocs.io/en/latest/}} package~\cite{Gorski_2005,Zonca2019}. We acknowledge the use of data from the Legacy Archive for Microwave Background Data Analysis (LAMBDA), part of the High Energy Astrophysics Science Archive Center (HEASARC). HEASARC/LAMBDA is a service of the Astrophysics Science Division at the NASA Goddard Space Flight Center. Part of the results presented here are based on observations obtained with \planck\ an ESA science mission with instruments and contributions directly funded by ESA Member States, NASA, and Canada. We also acknowledge that, this research used the resources of the National Energy Research Scientific Computing Center (NERSC), a U.S. Department of Energy Office of Science User Facility operated under Contract No. DE-AC02-05CH11231. Further, this work also made use of \texttt{SciPy}\footnote{\url{https://scipy.org/}}~\cite{scipy2020}, \texttt{NumPy}\footnote{\url{https://numpy.org/}}~\cite{Harris_2020}, \texttt{Astropy}\footnote{\url{https://www.astropy.org/}}~\cite{Astropy2013,Astropy2018,Astropy2022},
\texttt{Numba}\footnote{\url{https://numba.pydata.org/}}~\cite{Lam2015NumbaAL} and
\texttt{matplotlib}\footnote{\url{https://matplotlib.org/}}~\cite{Hunter2007}.

%REFERENCES
\bibliographystyle{unsrt}
\bibliography{ref_skp_proj3_new}

%\bsp	% typesetting comment
\end{document}